# Symmetry of ferroelectric switching and domain walls in hafnium dioxide

Guo-Dong Zhao,[1] Xingen Liu,[2] Wei Ren,[3] Xiaona Zhu,[1,*] and Shaofeng Yu.[1,4]

[1]*School of Microelectronics, Fudan University, Shanghai 200433, China*
[2]*School of Mathematical Information, Shaoxing University, Shaoxing 312000, China*
[3]*Physics Department, Shanghai Key Laboratory of High Temperature Superconductors, State Key Laboratory of Advanced Special Steel, International Centre of Quantum and Molecular Structures, Shanghai University, Shanghai 200444, China*
[4]*National Integrated Circuit Innovation Center, Shanghai 201204, China*

Email: xiaona_zhu@fudan.edu.cn

Hafnium dioxide ($HfO_2$) is a promising ferroelectric (FE) material for achieving high-density nonvolatile memory and neuromorphic computing, due to its compatibility with the mainstream integrated circuit technology and the surprisingly enhanced ferroelectricity by reduced thickness. The FE switching dynamics is essential to the device performance, but the complexity of $HfO_2$ atomic structure causes unknown of various FE switching paths and domain wall configurations. Here, we demonstrate that its low-barrier paths and domain walls can be comprehensively found and understood from a perspective of topological symmetry. By discussing pseudo-chirality and equivalent transformation relations in crystal with first principles and lattice modes, we classify and analyze 4 low-barrier FE switching paths and 93 irreducible topology domain wall configurations in $HfO_2$. Anisotropic switching mechanism is found based on the mobility investigation for 12 types of 180° side domain walls. This methodology is expected to be generally applicable to displacive ferroelectrics with low unit cell point group symmetries, and lay a foundation for mechanism study of the switching dynamics.

Ferroelectric (FE) materials have polar space group symmetry and non-zero net electric dipoles, which are switchable by the applied electric field. The hafnium dioxide ($HfO_2$) is drawing much attention as an emerging ferroelectric material. Since amorphous $HfO_2$ has been widely used as high-K dielectric in the modern complementary metal-oxide-semiconductor (CMOS) technology, the compatibility of FE $HfO_2$ with the mainstream integrated circuit technology makes it an attractive and promising material candidate in high-density nonvolatile memory, neuromorphic computing, and spintronics [1-4]. A generally studied ferroelectric phase of $HfO_2$ is the orthorhombic structure of $Pca2_1$ space group (OIII). The non-polar $P4_2/nmc$ tetragonal (T) phase is often considered as its parent phase [5] with less distortions from the high symmetry $Fm\bar{3}m$ cubic (C) phase. There are also possible existences of ferroelectric orthorhombic $Pmn2_1$ (OIV) phase [5,6] or rhombohedral phases [7].

Switching dynamics is an essential problem in the study of intrinsic FE property in $HfO_2$ [1,5,8-11] since it defines the operating voltage and dynamical performance of devices. Up to now, many efforts have been made to theoretically find possible FE switching paths [9,10,12] and domain wall (DW) configurations [8], but a thorough understanding has not yet been reached. This fact makes it hard to proceed to reliable large-scale mesoscopic simulations. Here with first principles and lattice mode analyses, we introduce a conceptual symmetry perspective that would naturally lead to a full view of basic structures, FE bulk switching mechanism, and DW configurations in OIII-$HfO_2$, and even in other complex ferroelectrics.

The T- and OIII-phase $HfO_2$ can be regarded as sequentially distorted from high symmetric C-Phase [5], and the distortions are described by symmetry-adapted modes. The highest parent C-phase structure has a compound lattice, and its Bravais lattice is face-centered cubic. It is known that from the C-phase, T-phase appears after freezing the unstable zone-boundary antipolar $X_2^-$ mode. Then, the OIII-phase appears mainly due to a further combining effect of soft zone-center polar $\Gamma_4^-$ and the zone-boundary antipolar $X_5^+$ modes, [1,12] accompanied by $X_3^-$ and $X_5^-$ modes with comparably small amplitudes (as shown in the Section S2 of Supplemental Materials [13]).

Owing to its typical non-symmorphic space group symmetry $Pca2_1$, OIII-$HfO_2$ crystal has a complex atomic structure: Except for the identity E, it only has screw ($2_1$) and glide (*c* and *a*) symmetries, without any rotation or mirror symmetries. So, the unit cell of $HfO_2$ has a chiral point group $C_1$. To study relative atomic positions and movements, we constrain our way of defining unit cells to a standard of Hf-cornered face-centered framework. Then, reflection operations orthogonal to coordinate axes can and only can generate eight pseudo-chiral configurations as shown in Fig. 1(a), which are labeled by hands. The term "pseudo" comes from the fact that the periodicity in crystal can superimpose the atomic structures among $S_{1\text{-}4}$ or $S_{1'\text{-}4'}$. The pseudo-chirality of the cells will be changed by improper rotation operations, i.e., reflection and inversion. We mark the inversion of a mode distortion direction as an inversion of "*sign*". It is clear from Fig. 1(b) that the operation $\hat{\sigma}_{ab}$ will inverse the *signs* of $\Gamma_4^-$, $X_{5x,z}^+$, and $X_2^-$ at the same time; the operation $\hat{\sigma}_{bc}$ inverses $X_2^-$ and $X_{5x,y}^{\pm}$; the operation $\hat{\sigma}_{ac}$ inverses $X_2^-$, $X_{5y,z}^+$, and $X_3^-$.

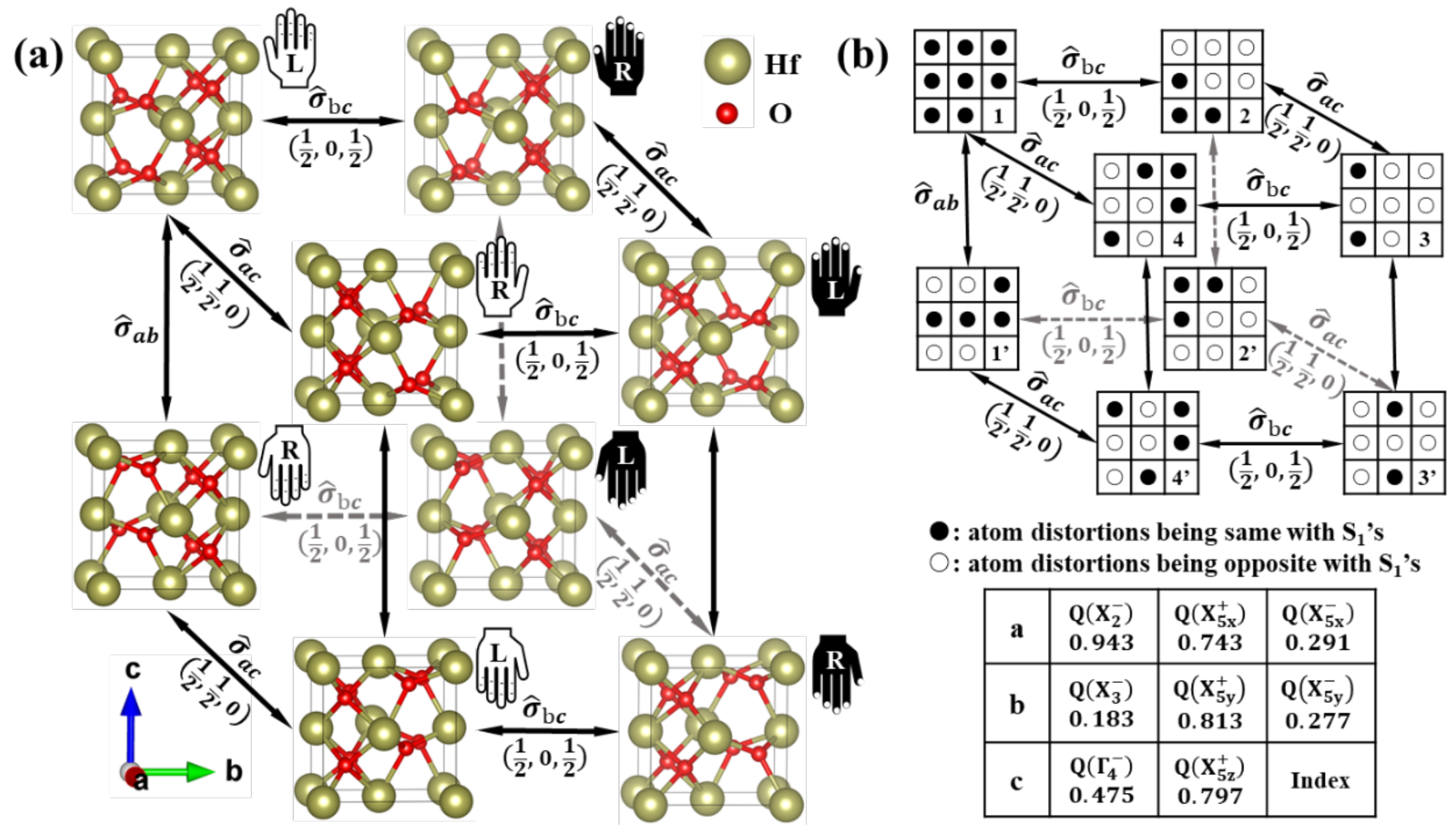

FIG. 1. (a) The eight kinds of pseudo-chiral $Pca2_1$ $HfO_2$ unit cells are characterized by atomic structures and their linking transformation operations. The hands point to the polarization directions, L (R) letters are denoted by the left (right) hands, and white (black) colors indicate the front (back) sides of hands. (b) The 3×3 chessboard representations abstract these 8 unit cells by their relative distortion directions of symmetry-adapted modes. The corresponding legend table shown below is understood as follows: "●" and "○" dots in the first, second, and third rows denote the relative distortion directions in *a*, *b*, and *c* axes, respectively, i.e., solid and open dots mean opposite distortion directions; the mode amplitudes are shown in the legend and are independent with their relative distortion directions; numbers 1~4 and 1'~4' in the lower right grid denotes the corresponding index numbers of unit cell structures $S_1$~$S_4$ and $S_{1'}$~$S_{4'}$ with upward and downward polarizations, respectively.

The adjacent transformation relations (reflection operations on three coordinate planes) among the 8 cells are shown above the arrows in Fig. 1. It is known that two orthogonal reflection operations lead to a 180° rotation operation (corresponding to a face diagonal connection in the big cube of Fig. 1), and three lead to an inversion operation (body diagonal in the big cube of Fig. 1), e.g., $\hat{\sigma}_{bc} \times \hat{\sigma}_{ac} = \hat{C}_{2c}$, and $\hat{\sigma}_{ab} \times \hat{\sigma}_{bc} \times \hat{\sigma}_{ac} = \hat{\imath}$; three-dimensional periodicity and the face-centered cubic Bravais lattice of its parent C-phase introduce a set of double-tracked transformation relations within cells of the same polarization directions, e.g., $\hat{\sigma}_{bc} = (1/2, 0, 1/2)$, and $\hat{\sigma}_{bc} \times \hat{\sigma}_{ac} = \hat{C}_{2c} = (0, 1/2, 1/2)$. The face-, base-, and body-centered Bravais lattices of parent phase ferroelectrics will introduce different translation-transformation relations. Ignoring the distortions of Hf lattice, here the fractional translations are taken as zeros and half-integers for simplification. We notice that Choe et al. [14] and Chen et al. [15] proposed similar ways of treating OIII-$HfO_2$ unit cells, and their works can be good references when we show below how to comprehensively utilize the crystal symmetries.

Since any known path where oxygen atoms penetrate vertical or horizontal Hf-planes suffers a large energy penalty [9], it is enough to consider FE switching paths where the 8 oxygen atoms only move within their own octants. Under such recognition, those 8 cells shown in Fig. 1 naturally provide a global view of bulk FE switching paths in OIII-$HfO_2$: With 4 ↑ and 4 ↓ cells serving as the initial and final states, there are only 4 inequivalent paths (1 ↑ to 4 ↓) due to translation transformation operations, e.g., (1/2, 0, 1/2) × {$S_2 \rightarrow S_{1'}$} = {$S_1 \rightarrow S_{2'}$}. Thus, we will discuss 4 independent FE switching paths with only $S_1$ as the initial structure.

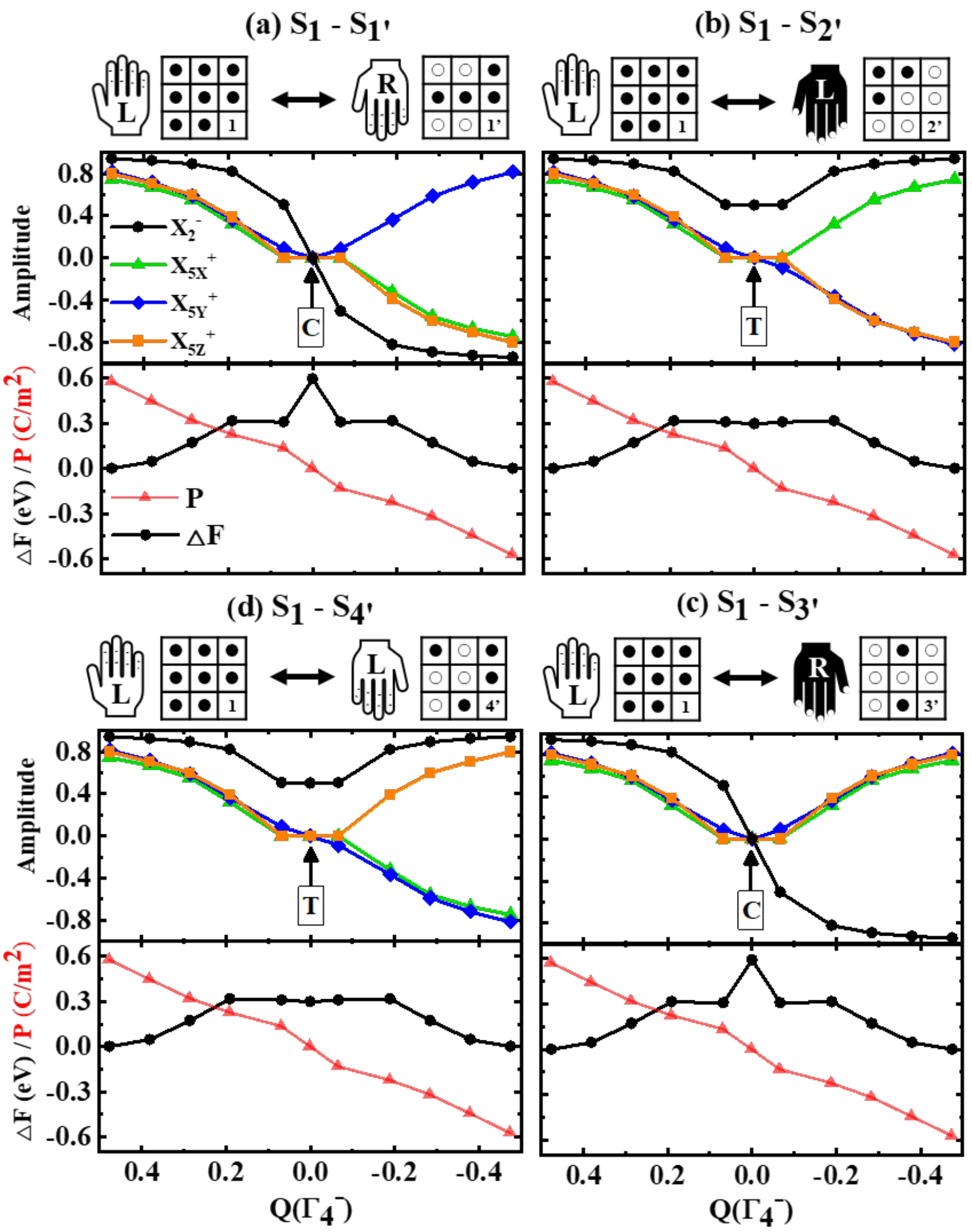

FIG. 2. The change of selected mode amplitudes, energies, and polarizations versus $\Gamma_4^-$ amplitude along the transition paths from $S_1$ to $S_{1'\sim4'}$. All the mode amplitudes of $S_1$ structure, whose *signs* are the same with "●" dots, are defined as positive.

To give an insight of FE-OIII $HfO_2$ transition paths, linear interpolated image structures between $S_1$ and $S_{1'\sim4'}$ structures are relaxed with fixed polar $\Gamma_4^-$ mode amplitudes [16], as shown in Fig. 2. This procedure gives the minimum energy path (MEP). During the switching, $HfO_2$ structure undergoes sequentially a symmetry transition, and its space group is: $Pca2_1$ - $Aba2$ - $P4_2/nmc$ (or $Fm\bar{3}m$) - $Aba2$ - $Pca2_1$. The C-phase ($Fm\bar{3}m$) reference structure doesn't have any distortions; the T-phase ($P4_2/nmc$) structure has only one non-zero amplitude of $X_2^-$ mode w.r.t. the reference C-phase; the $Aba2$ structure [12] has none-zero amplitude modes of $\Gamma_4^-$, $X_2^-$ and $X_5^+$ (only $y$-component $X_{5y}^+$). From Fig. 2, it is easy to find one of the C/T-phases must be traversed during the FE switching process. $X_2^-$ is the only mode which may not pass zero amplitude, only when the T-phase is traversed in $\{S_2 \rightarrow S_{2',4'}\}$. Whether $X_2^-$ passes zero makes a difference in the highest symmetry intermediate phase in paths. This is in correspondence with a recent report of Qi et al. [17].

The pseudo-chirality in $\{S_1 \rightarrow S_{1',3'}\}$ changes between the initial and final state structures, as shown in Fig. 2(a, c). Now periodicity is not involved in the standard frame of unit cell, so the pseudo-chirality turns to be a true chirality, and topological phase changes happen in $\{S_1 \rightarrow S_{1',3'}\}$. The topological transitions coincide with the traverse of all-zero mode amplitudes (C-phase), and therefore a crossing of higher energy barrier. As shown in lower panels of Fig. 2(b, d), the MEPs of $\{S_1 \rightarrow S_{2',4'}\}$ both have a barrier height of 315 meV/u.c., where the metastable T-phase constitutes a shallow concave at $Q(\Gamma_4^-) = 0$. While the MEPs of $\{S_1 \rightarrow S_{1',3'}\}$, shown in Fig. 2(a, c), have a higher barrier height of 594 meV/u.c., where C-phase is traversed and presents an energy peak. According to the above data and discussions, improper rotation operations $\hat{\sigma}_{ab}$ or $\hat{\sigma}_{ab} \times \hat{\sigma}_{bc} \times \hat{\sigma}_{ac} = \hat{\imath}$ inverse the *sign* of $X_2^-$. As for the comparison of paths $\{S_1 \rightarrow S_{2',4'}\}$ (or $\{S_1 \rightarrow S_{1',3'}\}$) with the same barrier heights, their zero-passing modes $X_{5x,z}^+$, $X_5^-$, and $X_3^-$ are opposite after traversing the highest symmetry structure, corresponding to a role exchange between the so-called "inert" and "active" oxygen layers [9,11]. Noteworthy, the *sign* of mode amplitude $X_{5y}^+$ must be inversed in energy-favored paths $\{S_1 \rightarrow S_{2',4'}\}$. Therefore, the anti-ferroelectricity in [010] is intercorrelated with the ferroelectricity in [001], and it could be inversed by an electric field, which is similar with two-dimensional ferroelectric material $In_2Se_3$ [18].

The DW configurations of such a complex pseudo-chiral $HfO_2$ structure has been ambiguous to be recognized and classified. Here we must introduce a general definition of orthogonal DWs as exampled in Fig. 3(a): We first determine the index of a unit cell on one DW side, then count integer numbers of unit cell length to a unit cell on the other side, and determine its index. The orthogonal DW interfaces are parallel with one of (100), (010), and (001) crystal planes. This procedure bypasses the relaxation of atomic details in DW, and allows us to conceptionally classify possible irreducible topology DW configurations. As shown in Fig. 3, by simultaneously shifting our frame of determining unit cell indexes on both DW sides, the DW configurations could be equivalently transformed. Thus, we can always find a $S_1$ cell on one side of DW in any configuration and take it as the reference domain, just like what we do in the learning of FE switching paths.

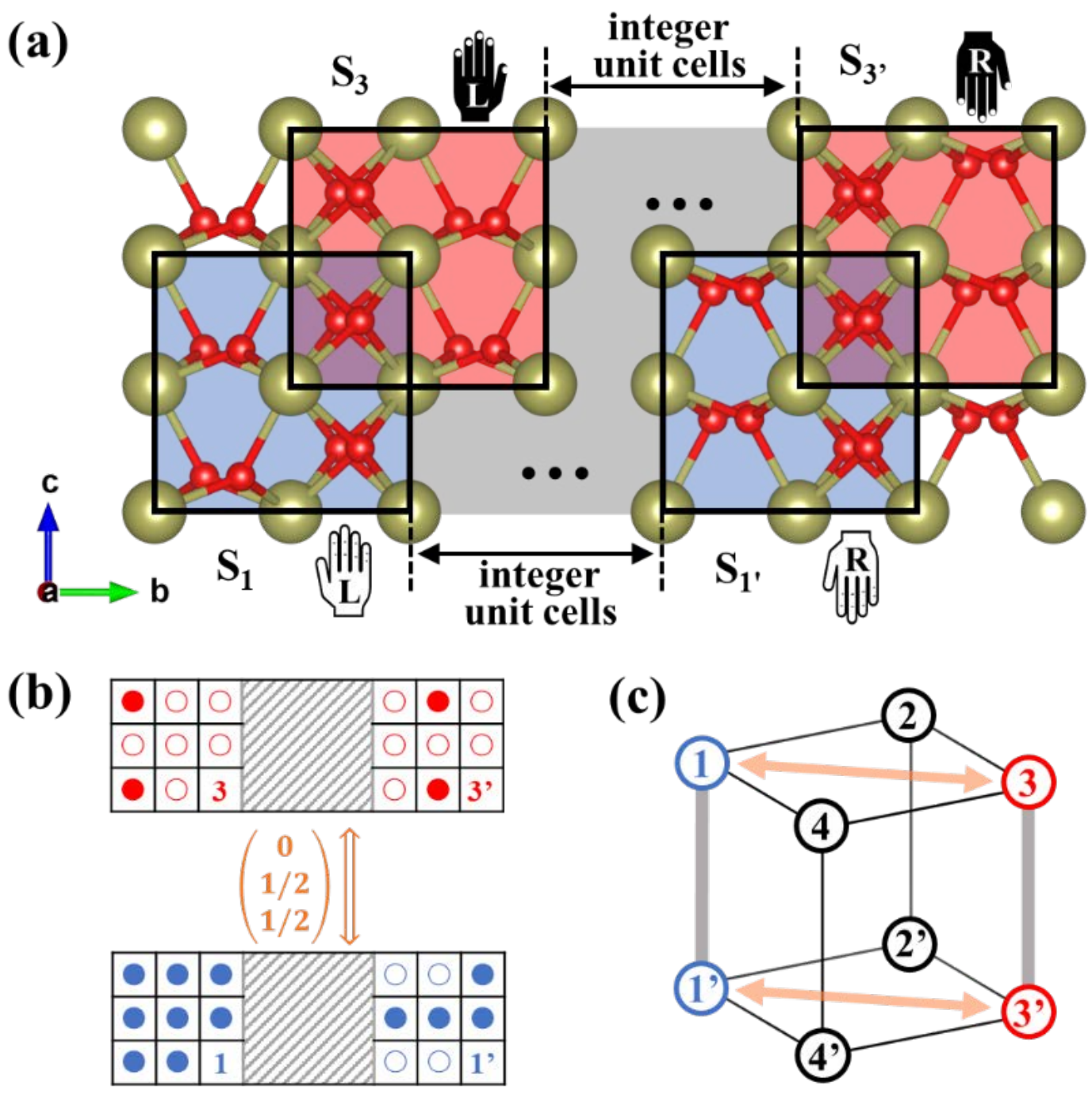

FIG. 3. (a) Two equivalent sets of 180° shoulder-to-shoulder DWs $S_1$-$S_{1'}$ and $S_3$-$S_{3'}$ shown in atomic structures. The grey area is a general definition of DW, within which *signs* of specific modes are turned, despite unknown atomic details. (b) The abstracted chessboards notion of equivalent DWs, showing the equivalence in modes. (c) A sketched of DWs and their transformation relation in the index cube.

Based on the general definition, different types of orthogonal DWs in *Pca*$2_1$-$HfO_2$ can be uniformly named as:

$$D_1 D_2^{\pm} I^{(\prime)}, \tag{1}$$

where $D_1$ indicates the reference domain, and $D_2$ the second domain; D = S (Shoulder, *ac*-plane), F (Face, *bc*-plane), H (Head, *ab*-plane on the positive side of polarization) or T (Tail, *ab*-plane on the negative side of polarization), referring to the interface of reference and the second domain on either side of DW; the superscript "±" of $D_2$ suggests its normal (+) or reverse (-) lattice-vector correspondence order to the reference domain, e.g., a DW with an interface "*ab* | *bc*", where orderly $a_{ref} \parallel b_{2nd}$ and $b_{ref} \parallel c_{2nd}$, can be marked as $HF^{+}I^{(\prime)}$ or $TF^{+}I^{(\prime)}$, while opposite ordered "*ab* | *cb*" should be as $HF^{-}I^{(\prime)}$ or $TF^{-}I^{(\prime)}$; $I^{(\prime)}$ = 1, 1', ...4, 4', denoting the unit cell index of the second domain. Combining the translation transformation operations described in Fig. 3 and rotation transformation operations, one can deduce 93 irreducible types of inequivalent orthogonal DWs in ferroelectric *Pca*$2_1$-$HfO_2$: 15 types of 0° DWs, 24 types of 180° DWs, and 54 types of 90° DWs. This can be regarded as an extension of the classification of Ref. [8]. All the irreducible DW configurations are concluded in Table 1, and more detailed sketches are in Supplemental Materials [13] for further interests.

TABLE 1. Concluding the numbers of different domain wall configurations. Note one unique DW configuration could have multiple names, and we only choose one to show.

| **$D_1D_2$** | + | - | **Total** = 96 (-3) |
|---|---|---|---|
| SS | 8 (-1) | 6 | 14 (-1) |
| FF | 8 (-1) | 8 | 16 (-1) |
| FS | 8 | 8 | 16 |
| HT | 4 (-1) | 2 | 6 (-1) |
| HH | 4 | 2 | 6 |
| TT | 4 | 2 | 6 |
| HS | 4 | 4 | 8 |
| TS | 4 | 4 | 8 |
| HF | 4 | 4 | 8 |
| TF | 4 | 4 | 8 |

To link from classified DW configurations to DW mobilities, here we exhibit the DW propagation energy barriers of 12 types 180° side DWs ($SS^+I'$, $FF^+I'$, and $FS^+I'$), whose interfaces are parallel with polarization directions of both domains. 180° DWs have drawn many attentions [1,17], and they dominant the switching dynamics in thin films [19]. Equally spaced two DWs are constructed in 1 × 8 × 1 supercells, i.e., with an interval of 4 unit cells length. The lattice constants and atomic coordinates of multi-domain structures are fully relaxed. DW energy is defined as $E_{DW} = (E_{MD} - E_{SD}) / 2S$, where $E_{MD}$ is the total energy of a multi-domain supercell, $E_{SD}$ represents the total energy of a single-domain supercell, and S denotes the interface area of DW in a multi-domain supercell. The DW propagation energy is an important intrinsic factor to determine DW mobility, and it is calculated by moving one DW for one unit cell length. The corresponding MEPs are plotted in Fig. 4, and the results are concluded in Table 2.

TABLE 2. DW energies ($E_{DW}$) and DW propagation barriers (ΔE) of 180° side DWs in OIII-$HfO_2$.

| DW type | $E_{DW}$ | | ΔE |
|---|---|---|---|
| | (mJ/m$^2$) | (meV/Å$^2$) | (meV) |
| $SS^+1'$ | -38 | -2 | 972 |
| $SS^+2'$ | 448 | 28 | 79 |
| $SS^+3'$ | 386 | 24 | 693 |
| $SS^+4'$ | 399 | 25 | 199 |
| $FF^+1'$ | 293 | 18 | 1148 |
| $FF^+2'$ | 255 | 16 | 219 |
| $FF^+3'$ | 809 | 50 | 186 |
| $FF^+4'$ | 506 | 32 | 23 |
| $FS^+1'$ | 641 | 40 | 72 |
| $FS^+2'$ | 645 | 40 | 105 |
| $FS^+3'$ | 641 | 40 | 71 |
| $FS^+4'$ | 641 | 40 | 104 |

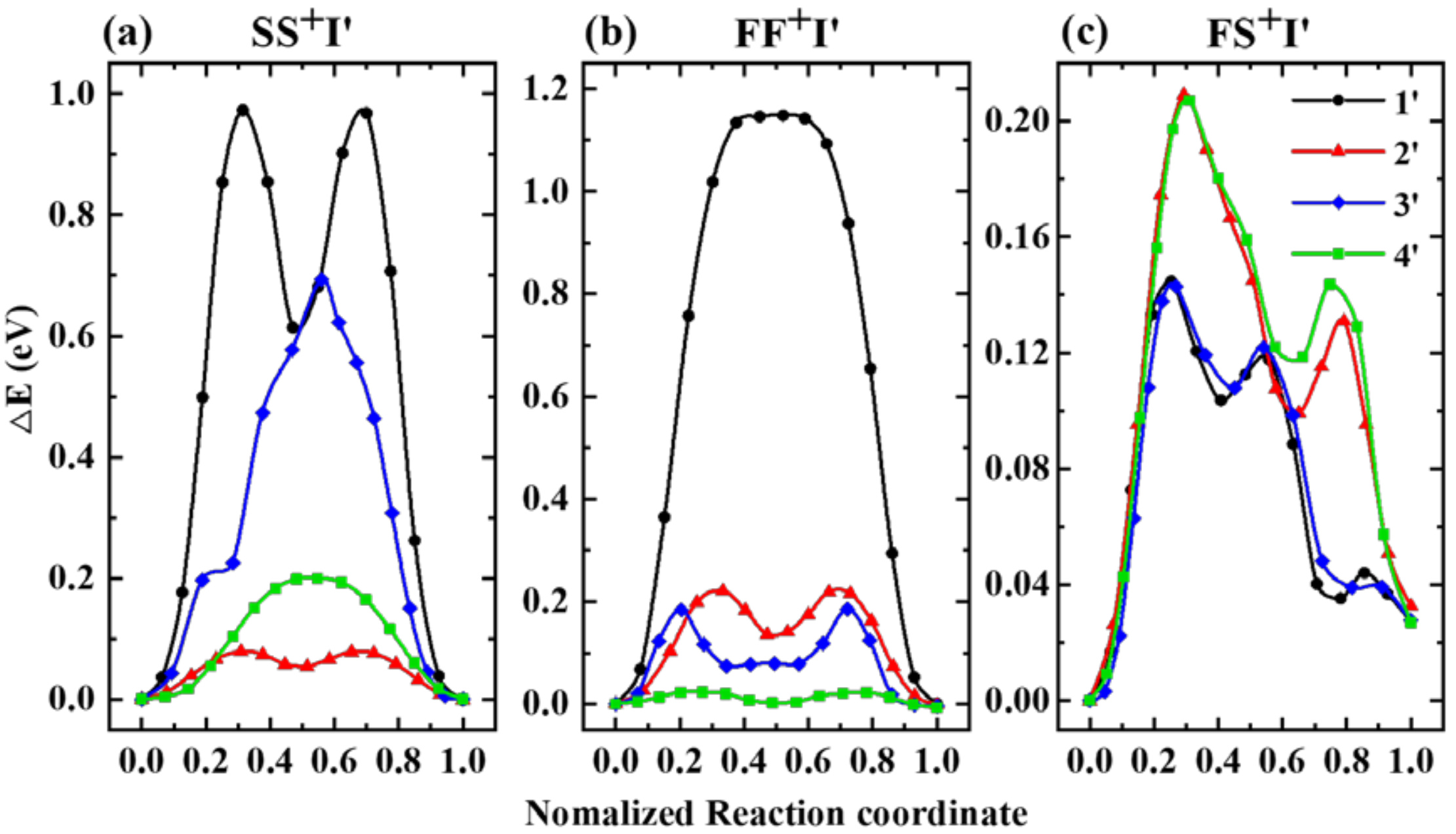

FIG. 4. The minimum energy paths, of DW propagation for a length of 1 unit cell, are calculated in 180° side DWs $SS^{+}I'$, $FF^{+}I'$, and $FS^{+}I'$.

From Table 2, we first see the $E_{DW}$ of $SS^{+}I'$, $FF^{+}I'$, and $FS^{+}I'$ types of 180° side DWs differ: they span in the range of -38 ~ 448 mJ/m$^2$, 253 ~ 809 mJ/m$^2$, and 641 ~ 645 mJ/m$^2$, respectively. $FF^{+}1'$ and $FF^{+}2'$ have lower $E_{DW}$ than that of $SS^{+}I'$ DWs, except $SS^{+}1'$. The well-known negative $E_{DW}$ of $SS^{+}1'$ is due to the emergence of lower energy of *Pbca* anti-polar phase [1,20] at the DW area. $FS^{+}I'$ type 180° DWs have an averagely higher $E_{DW}$, since they own a larger lattice mismatch between two adjacent domains. The DW propagation energy barriers (ΔE) are also listed in Table 2. ΔE of $SS^{+}I'$, $FF^{+}I'$, $FS^{+}I'$ types 180° DWs ranges in 79 ~ 972 meV, 23 ~ 1148 meV, and 71 ~ 105 meV, respectively. The ranks of barriers ΔE are not positively correlated with that of $E_{DW}$, and the ΔE of $FF^{+}I'$ DWs are not necessarily lower than that of $SS^{+}I'$ type 180° DWs. Obviously, anisotropic nucleation rates with different critical nucleus sizes in [100] and [010], together with anisotropic growth rates are expected during switching of thin film OIII-$HfO_2$, different with prototype Perovskite ferroelectrics [19,21]. Moreover, The data in Table 2 validates the high activation field explanation of Lee et al. [1]. We notice that Choe et al. [14] compared $SS^{+}1'$ and $SS^{+}4'$, marked as (1;1) and (2;0) in their work, respectively. We have discrepancies in the absolute numbers of ΔE, which could be attributed to our relaxed lattice constants of DW structures and different exchange-correlation functionals. Detailed MEPs and atomic structure shots of typical images are provided in Supplemental Materials [13] for further interests.

In summary, the symmetry of complex ferroelectric OIII-$HfO_2$ structure is elucidated in the aspects of FE switching paths and DW configurations. By transformation relations and symmetry-adapted mode distortion directions, eight pseudo-chiral analog unit cells are defined. Then four bulk $HfO_2$ FE switching paths emerge naturally and can be decomposed w.r.t. the change of mode amplitudes, polarizations, and energies. Higher barriers during transitions are attributed to the topology phase change where the chirality is changed, or the inversion of $X_2^-$ distortion directions. The difference between energy-degenerate paths is described as the inversion of $X_{5x,z}^+$, $X_5^-$, and $X_3^-$ distortions. We propose that by our general definition of DWs together with equivalent transformation relations, one can determine 93 mutations of irreducible topology DW configurations. Based on 180° side DW configurations, the possible existence of anisotropic DW lateral expansion is proposed. This work will help understand the symmetry of the OIII-$HfO_2$ structure, and the methodology is generally applicable to FE switching paths and irreducible topology DW configurations in other complex ferroelectric materials with low unit cell point group symmetries.

X. Zhu thanks the Shanghai Sailing Program. X. Liu gives special thanks to the Research Start-up Fund Project of Shaoxing University. W. Ren thanks the support by the National Natural Science Foundation of China (12074241, 11929401, 52130204), the Science and Technology Commission of Shanghai Municipality (19010500500, 20501130600), High Performance Computing Center, Shanghai University, and Key Research Project of Zhejiang Lab (No. 2021PE0AC02). G.D. Zhao thanks Fanhao Jia and Chang Liu from ICQMS of Shanghai University for valuable discussions.

The supplemental materials of
Symmetry of ferroelectric switching and domain walls in hafnium dioxide

Guo-Dong Zhao,[1] Xingen Liu,[2] Wei Ren,[3] Xiaona Zhu,[1,]* Shaofeng Yu[1,4]

[1]*School of Microelectronics, Fudan University, Shanghai 200433, China*
[2]*School of mathematical information, Shaoxing University, Shaoxing 312000, China*
[3]*Physics Department, Shanghai Key Laboratory of High Temperature Superconductors, State Key Laboratory of Advanced Special Steel, International Centre of Quantum and Molecular Structures, Shanghai University, Shanghai 200444, China*
[4]*National Integrated Circuit Innovation Center, Shanghai 201204, China*

xiaona_zhu@fudan.edu.cn

**Section S1.** Calculation methods

All first principles calculations in this work are based on the density functional theory (DFT) with local-density-approximation (LDA) exchange-correlation functional [1].

Data in Fig. 1 and Fig. 2 are calculated with the ABINIT [2,3] package, using the Projected-augmented wave (PAW) [4,5] pseudopotentials from the Jollet–Torrent–Holzwarth (JTH) projector augmented-wave atomic dataset library [6] (12 valence electrons for Hf, and 6 for O element). The cutoff energy is 50 Hartree, relaxations are converged when the force on each atom is below $1\times10^{-5}$ Hartree/Bohr, with a 4 × 4 × 4 Monkhorst-Pack *k*-point grid. The constraining of mode amplitudes follows the method described in Ref. [7] (the "wtatcon" parameter). Symmetry-adapted modes of optimized structures are decomposed via the AMPLIMODES [8,9] module of the Bilbao Crystallographic Server [10-12]. The crystal structure visualizations are illustrated by the VESTA [13] software.

Data in Fig. 4 and Table 2 are based on the PAW method implemented in the efficient Vienna Ab initio Simulation Package (VASP) [14,15]. The cutoff energy for the plane-wave-basis set is 600 eV. We included 10 valence electrons for Hf element, and 6 for O element in the pseudopotentials. For the structures of unit cell and DWs: lattice constants and atomic coordinates are fully relaxed until the maximum force on ions being less than 0.001 eV/Å and 0.01 eV/Å, respectively; the distances between nearing k-points in the Γ-centered Monkhorst Pack k-grids are set to be smaller than 0.1 $Å^{-1}$ and 0.3 $Å^{-1}$, respectively. The DW propagation barriers are calculated via the climbing image nudged elastic band (CI-NEB) [16] method.

From VASP, the relaxed lattice constants of orthorhombic $Pca2_1$-$HfO_2$ are $a$ = 5.162 Å, $b$ = 4.960 Å, and $c$ = 4.978 Å, almost identical with the results from ABINIT ($a$ = 5.161 Å, $b$ = 4.958 Å, $c$ = 4.978 Å). Macroscopic polarization of $Pca2_1$-$HfO_2$ is evaluated as 58 μC/cm$^2$ in the berry phase expression [17], and linear inserted transition images are considered to avoid quanta of polarization. This is similar with the result from ABINIT (57 μC/cm$^2$). Bulk switching energy barriers are 327 meV/u.c. for {$S_1$ → $S_{2',4'}$} paths, and 758 meV/u.c. for {$S_1 \rightarrow S_{1',3'}$} paths. The tetragonal phase transition barrier is similar with ABINIT's (315 meV/u.c.), while the barrier through cubic phase is relatively higher than the ABINIT result (594 meV/u.c.).

Two equally spaced DWs are constructed by building blocks from Fig. S2 in 1 × 8 × 1 supercells, i.e., with an interval of 4 unit cell lengths. The lattice constants and atomic coordinates of multi-domain structures are fully relaxed in a stepwise manner. DW energy is defined as $E_{DW} = (E_{MD} - E_{SD}) / 2S$, where $E_{MD}$ is the total energy of a multi-domain supercell, $E_{SD}$ represents the total energy of a single domain supercell, and S denotes the interface area of DW in a multi-domain supercell. The DW propagation energy is calculated by normally moving one DW for one unit cell length, i.e., expanding one domain from 4 to 5, and shrink the other to 3 unit cell lengths.

**Section S2.** Symmetry-adapted modes of *Pca*$2_1$-$HfO_2$

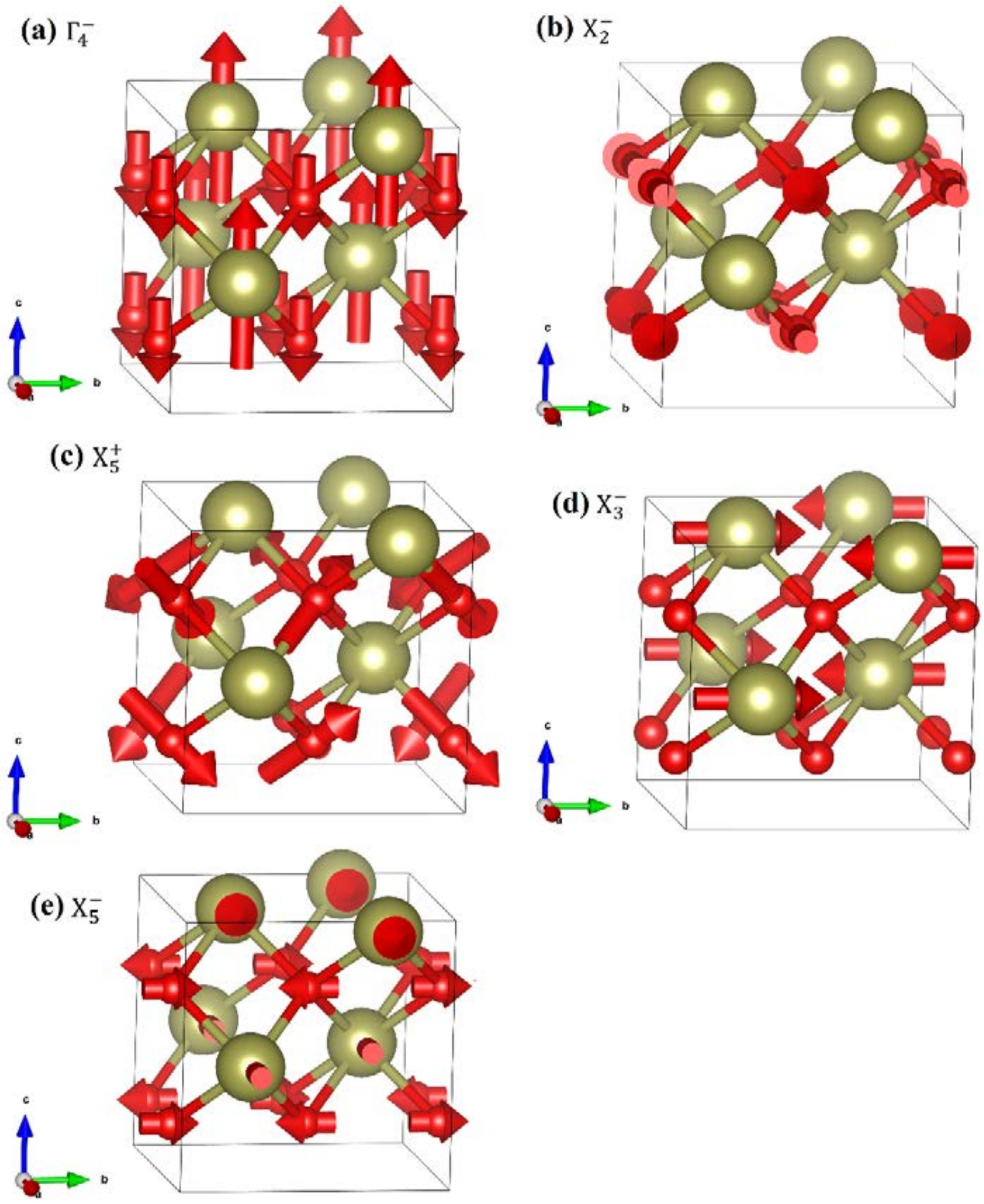

FIG. S1. The symmetry-adapted modes of Pca$2_1$ $HfO_2$ shown in atomic views: (a) $\Gamma_4^-$; (b) $X_2^-$; (c) $X_5^+$; (d) $X_3^-$; (e) $X_5^-$.

TABLE S1. Normalized basis symmetry modes:

$\Gamma(0, 0, 0)$

$\Gamma_4^-$: Hf1-1

| Atom | δx | δy | δz |
|---|---|---|---|
| Hf1 | 0 | 0 | 0.099586 |

$\Gamma_4^-$: O1-1

| Atom | δx | δy | δz |
|---|---|---|---|
| O1 | 0 | 0 | 0.070418 |
| O1_2 | 0 | 0 | 0.070418 |

X(0, 1, 0)

$X_5^+$: O1-1(z)

| Atom | δx | δy | δz |
|---|---|---|---|
| O1 | 0 | 0 | 0.070418 |
| O1_2 | 0 | 0 | -0.070418 |

$X_5^+$: O1-2(y)

| Atom | δx | δy | δz |
|---|---|---|---|
| O1 | 0 | -0.070418 | 0 |
| O1_2 | 0 | 0.070418 | 0 |

$X_5^+$: O1-3(x)

| Atom | δx | δy | δz |
|---|---|---|---|
| O1 | -0.070418 | 0 | 0 |
| O1_2 | -0.070418 | 0 | 0 |

$X_2^-$: O1-1(x)

| Atom | δx | δy | δz |
|---|---|---|---|
| O1 | 0.070418 | 0 | 0 |
| O1_2 | -0.070418 | 0 | 0 |

$X_3^-$: Hf1-1(y)

| Atom | δx | δy | δz |
|---|---|---|---|
| Hf1 | 0 | -0.099586 | 0 |

$X_5^-$: Hf1-1(x)

| Atom | δx | δy | δz |
|---|---|---|---|
| Hf1 | 0.099586 | 0 | 0 |

$X_5^-$: O1-2(y)

| Atom | δx | δy | δz |
|---|---|---|---|
| O1 | 0 | -0.070418 | 0 |
| O1_2 | 0 | -0.070418 | 0 |

TABLE S2. Normalized polarization vectors (displacements in cell relative units) in the asymmetric unit:

| | $\Gamma_4^-$ | | |
|---|---|---|---|
| Atom | δx | δy | δz |
| Hf1 | 0 | 0 | 0.0813 |
| O1 | 0 | 0 | -0.0407 |
| O1_2 | 0 | 0 | -0.0407 |
| | $X_5^+$ | | |
| Atom | δx | δy | δz |
| Hf1 | 0 | 0 | 0 |
| O1 | 0.0385 | 0.0421 | 0.0413 |
| O1_2 | 0.0385 | -0.0421 | -0.0413 |
| | $X_2^-$ | | |
| Atom | δx | δy | δz |
| Hf1 | 0 | 0 | 0 |
| O1 | -0.0704 | 0 | 0 |
| O1_2 | 0.0704 | 0 | 0 |
| | $X_3^-$ | | |
| Atom | δx | δy | δz |
| Hf1 | 0 | -0.0996 | 0 |
| O1 | 0 | 0 | 0 |
| O1_2 | 0 | 0 | 0 |
| | $X_5^-$ | | |
| Atom | δx | δy | δz |
| Hf1 | 0.0826 | 0 | 0 |
| O1 | 0 | -0.0394 | 0 |
| O1_2 | 0 | -0.0394 | 0 |

**Section S3.** All irreducible DW configurations of $Pca2_1$-$HfO_2$

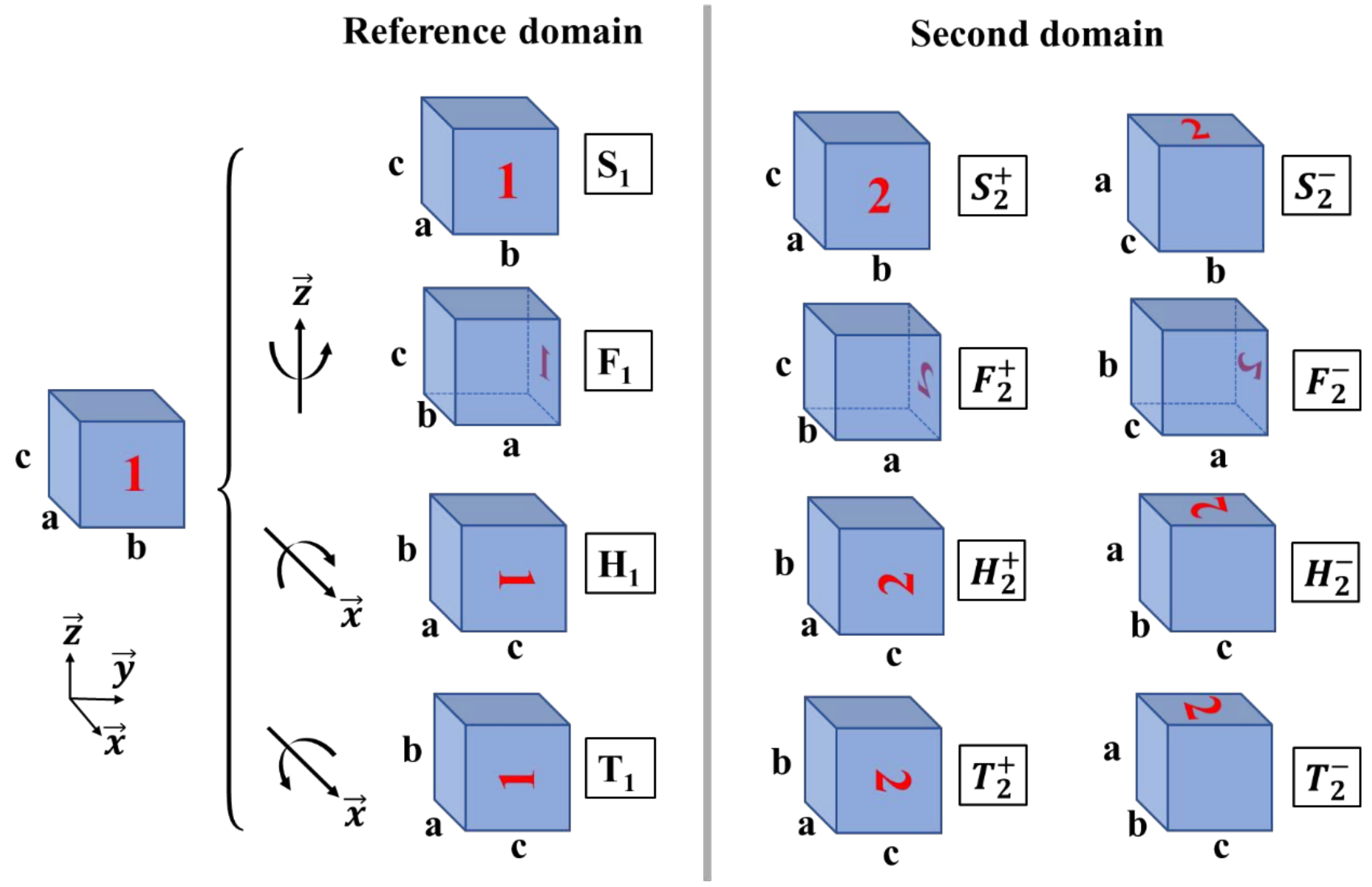

FIG. S2. A sketch diagram showing how do we define the domain steric relations to name the DW configurations. We should always put the DW plane parallel with *xz*-plane, by properly rotating the lattices of reference domain and the second domain.

The way of defining "±" and "index" of the second domain can be various. Here our specific way of defining unit cell indexes in an arbitary DW is shown in Fig. S2. A standard viewing angle is defined to compare and determine the indexes, according to the index cube of Fig. 1 in the main-text. By default, we put the reference domain (indexed as $S_1$) on the negative side of *y*-axis, and the second domain on the positive side of *y*-axis. In the full DW name, letter for reference domain will be written on the left. The subscript "1" is for the reference domain and "2" is for the second domain, and they will be shown when comparison is made between default and non-default DW settings.

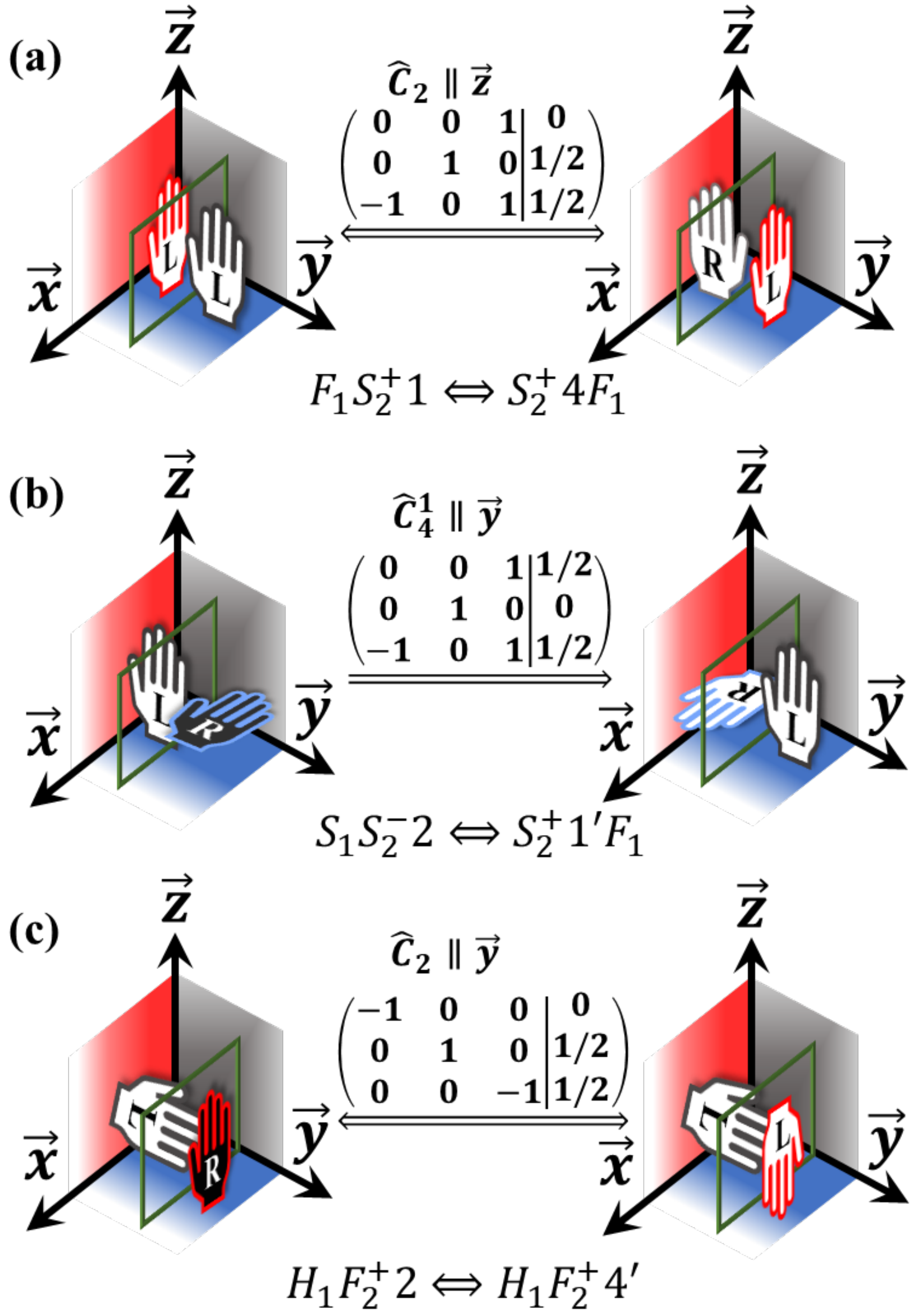

FIG. S3. The reducing effect of some combined equivalent relations with both rotation and translation, represented by rotation-translation operators $(\boldsymbol{R}|\boldsymbol{t}_{l_1 l_2 l_3})$. Example DW configurations: (a) 2 equivalent 0° face-to-shoulder DWs $F_1 S_2^+ 1 \Leftrightarrow S_2^+ 4 F_1$, (b) 2 equivalent 90° shoulder-to-shoulder DWs $S_1 S_2^- 2 \Leftrightarrow S_2^+ 1' F_1$, and (c) 2 equivalent 90° head-to-face DWs $H_1 F_2^+ 2 \Leftrightarrow H_1 F_2^+ 4'$. Hand legends denote two cells separated by an integer number of unit cell lengths on two sides of a DW interface, which is indicated as a green frame. The outline colors of hands suggest their parallel crystal surfaces (grey to *yz*-plane, red to *xz*-plane, and blue to *xy*-plane).

As shown in Fig. S3, apart from the translation equivalent relations shown in Fig. 3 of the main-text, there are multiple equivalent relations with both translations and rotations. Note that in a standard naming process, the rotation-translation operators $(\boldsymbol{R}|\boldsymbol{t}_{l_1 l_2 l_3})$ are all w.r.t. the coordinate system, not lattice vectors.

After some tedious deductions by three-dimensional imaginations, we conclude that there are totally 93 inequivalent DW configurations. Note that 3 types of single-domain configurations in FF, SS, and HT are reduced from 96. All the DW configurations and some typical equivalent relations are sketched in Fig. S4~S10. The number of DW configurations are concluded in Table 1 of the main-text, and a raw outline concluding degrees is given here:

- Orthorgonal DWs (93)
  - 0° (15)
    - FF (3), FS (4)
    - HT (5)
    - SS (3)
  - 180° (24)
    - FF (4), FS (4)
    - TT (6)
    - HH (6)
    - SS (4)
  - 90° (54)
    - FF (8), FS (8)
    - TS (8), TF (8)
    - HS (8), HF (8)
    - SS (6)

The lattice mismatch brought by the longer *a*-constant with the other two constants is ~ 4% in OIII-$HfO_2$, and it will increase the DW energy. To avoid this mismatch, the *bc*-surface (Face) of two domains should be mutually parallel, and this reduces the number of proper orthogonal DW configurations to 37.

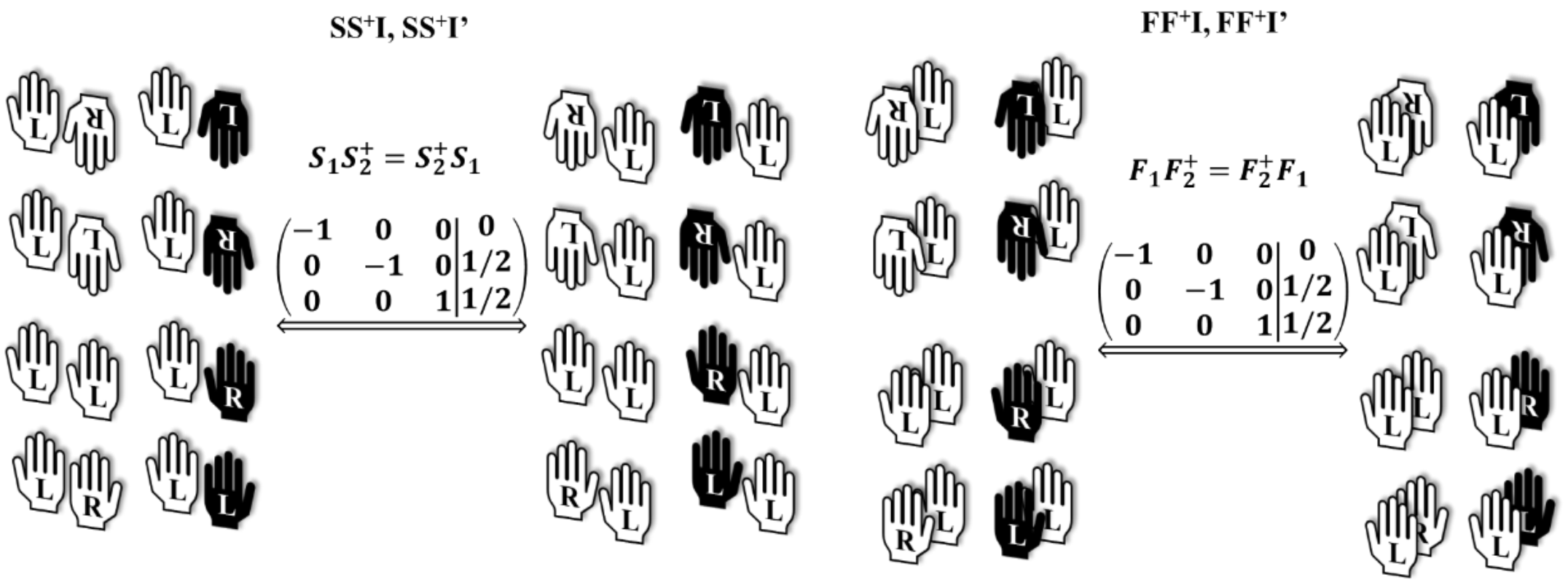

FIG. S4. Irreducible DW configurations: 8 types of $SS^+I$ and $SS^+I'$; 8 types of $FF^+I$ and $FF^+I'$.

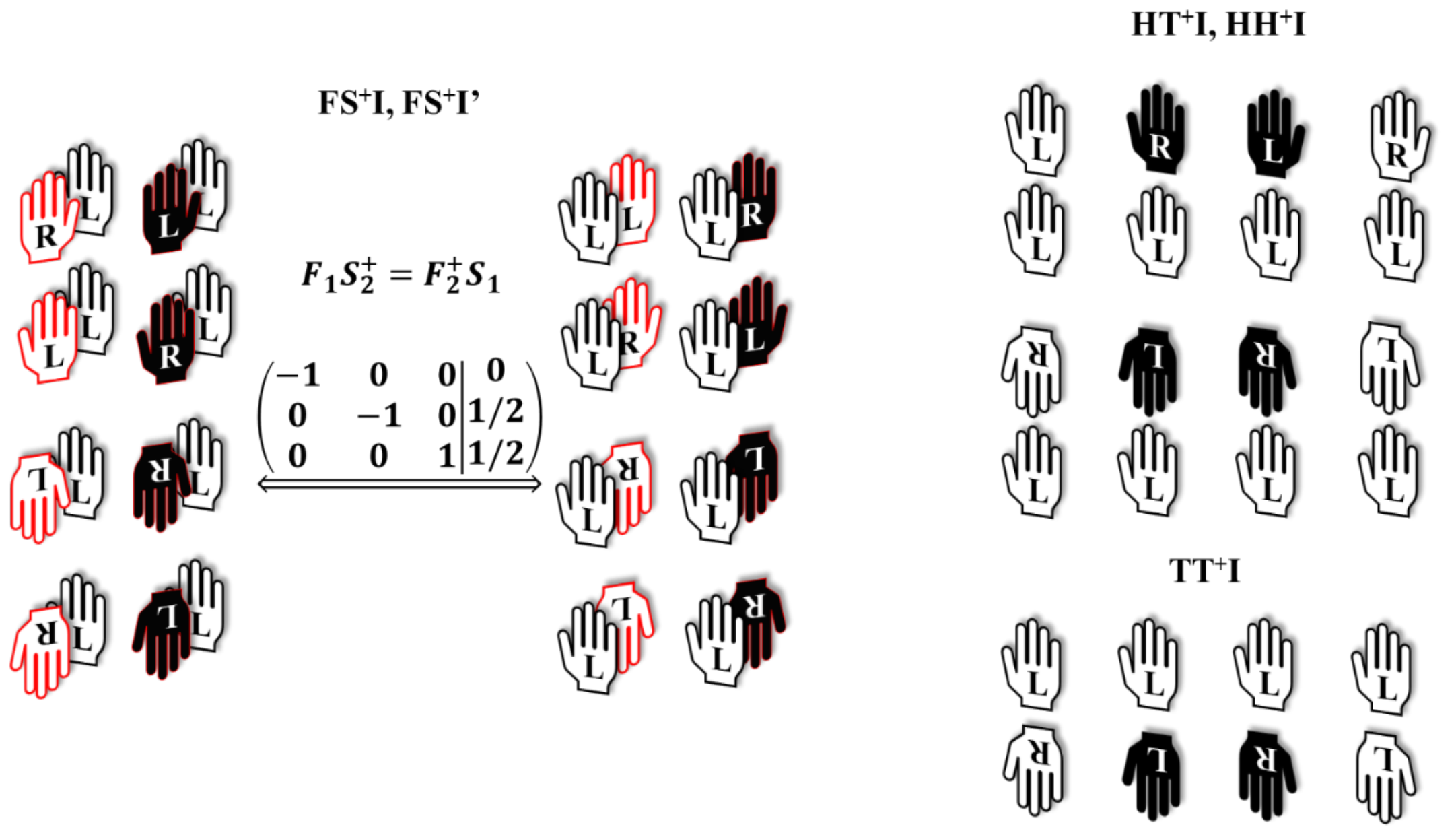

FIG. S5. Irreducible DW configurations: 8 types of FS$^+$I and FS$^+$I’; 8 types of HT$^+$I and HT$^+$I; 4 types of TT$^+$I’.

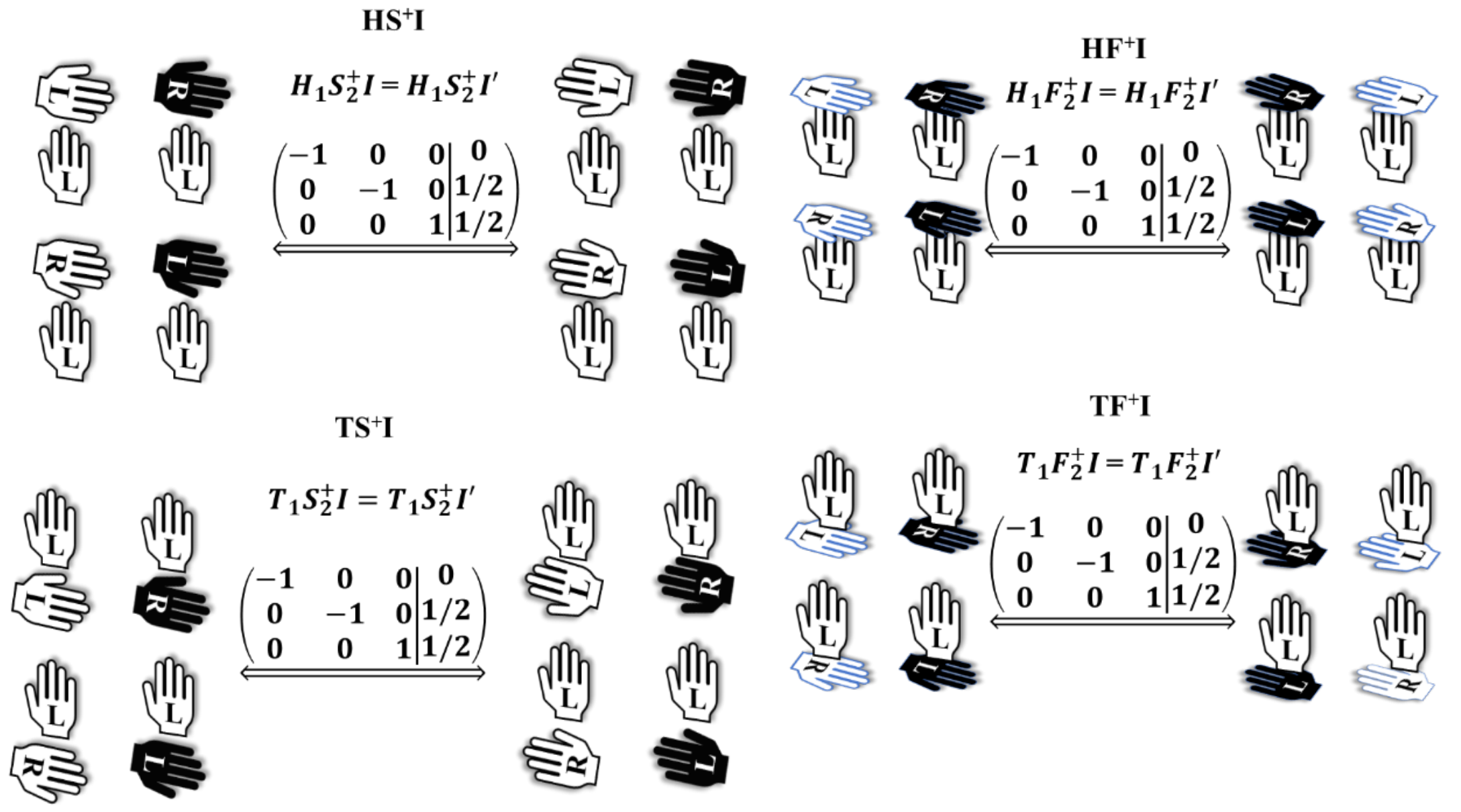

FIG. S6. Irreducible DW configurations: 4 types of HS$^+$I; 4 types of TS$^+$I; 4 types of HF$^+$I; 4 types of TF$^+$I’.

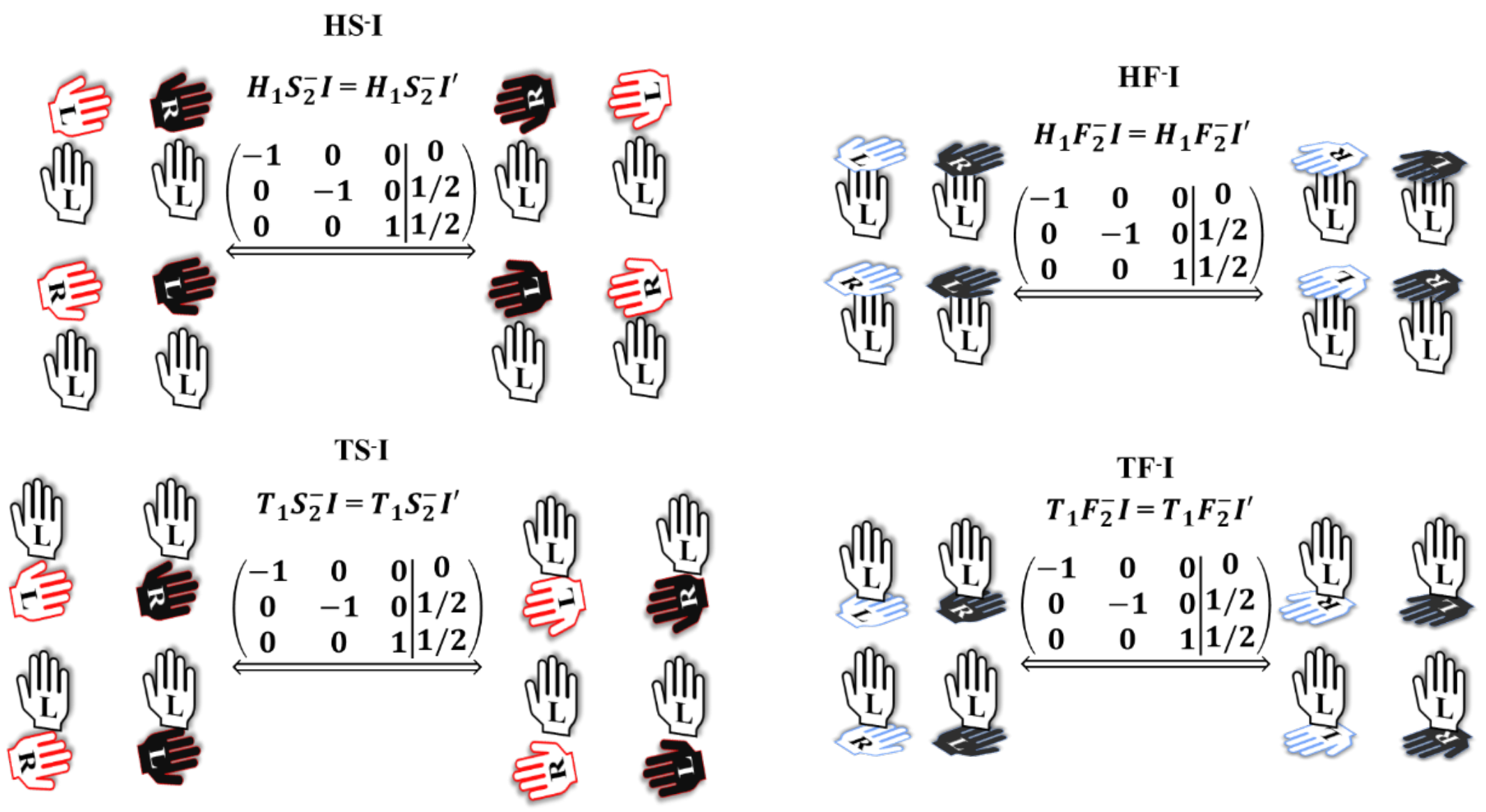

FIG. S7. Irreducible DW configurations: 4 types of HS$^-$I; 4 types of TS$^-$I; 4 types of HF$^-$I; 4 types of TF$^-$I'.

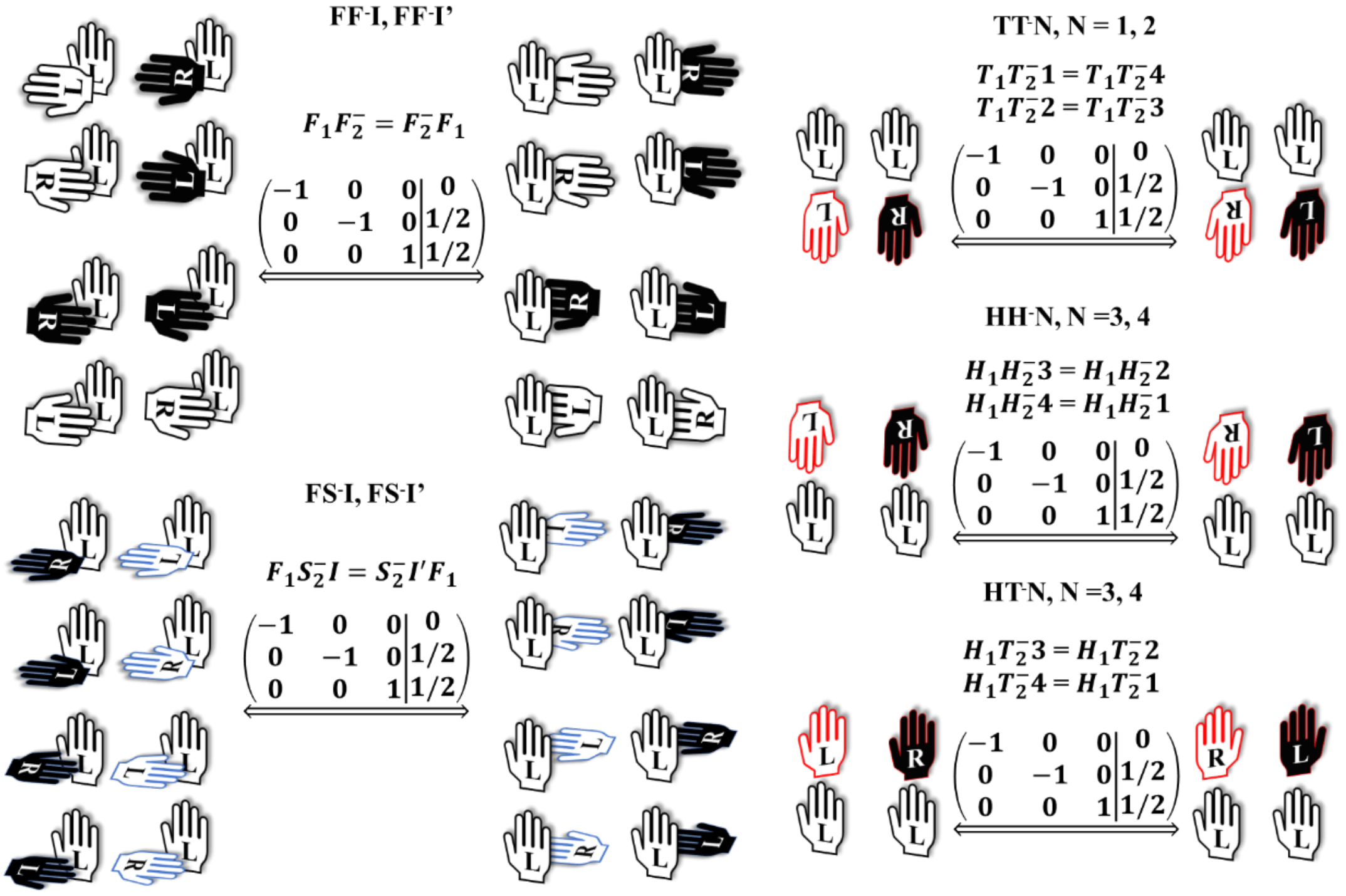

FIG. S8. Irreducible DW configurations: 8 types of FF$^-$I and FF-I'; 8 types of FS$^-$I and FS$^-$I'; 4 types of TT$^-$N, where N = 1, 2; 2 types of HH$^-$N, where N = 3, 4; 2 types of HT$^-$N, where N = 3, 4. Here N could be 1, 2, or also 3,4, making no difference.

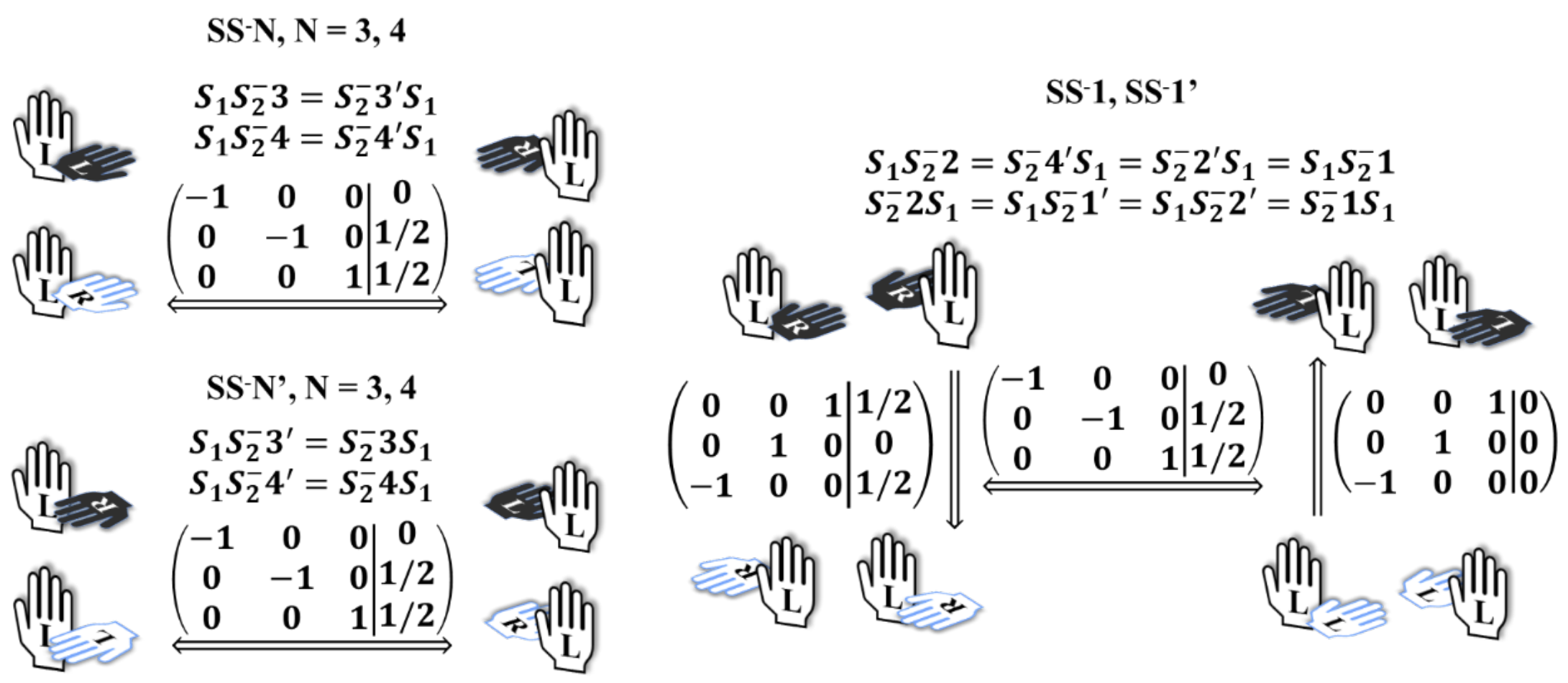

FIG. S9. Irreducible DW configurations: 4 types of SS-N and SS-N', where N = 3, 4; 2 types of SS-1 and SS-1'.

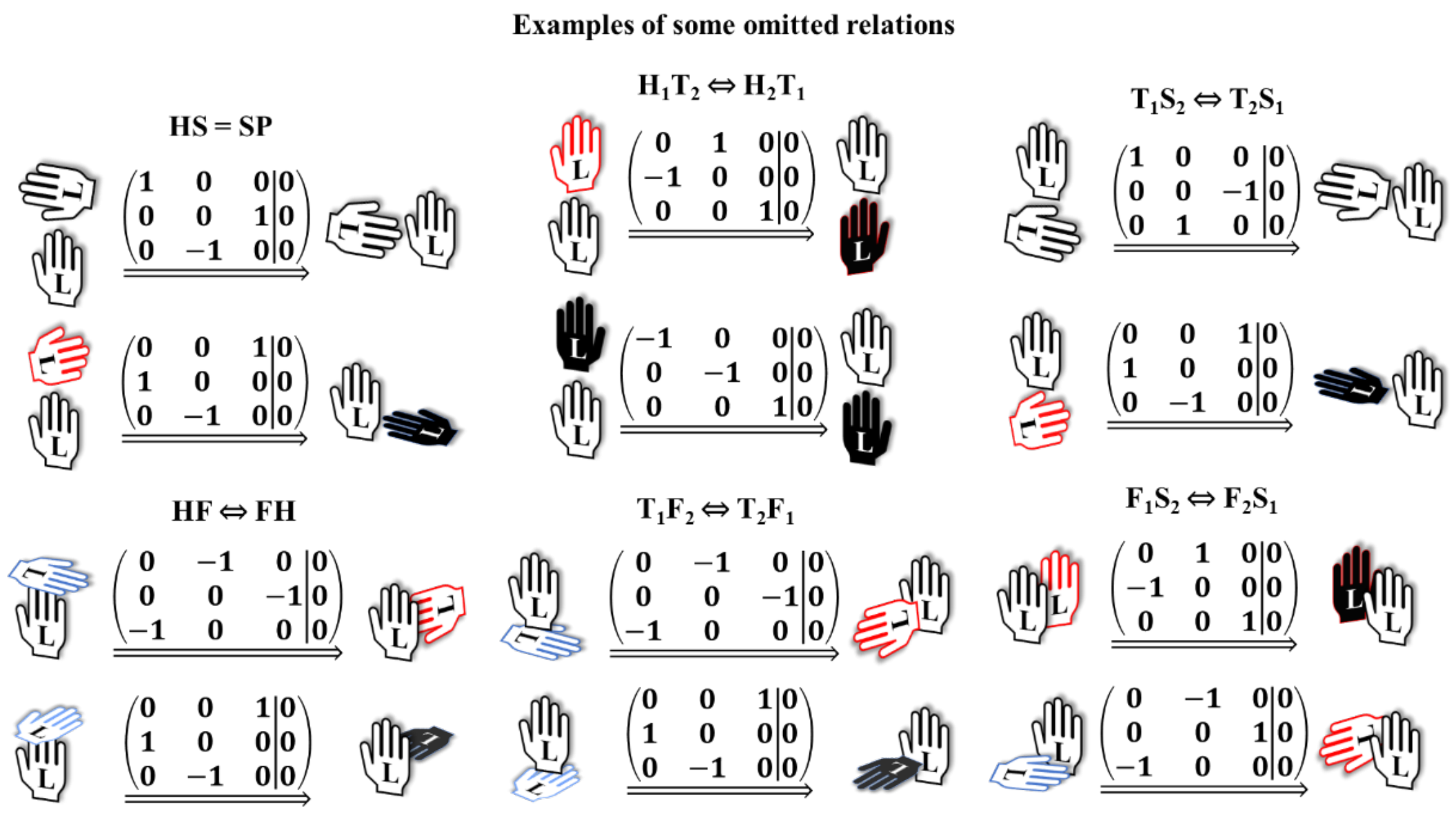

FIG. S10. Examples of some omitted equivalent relations.

**Section S4.** Structural details of DW propagations in 12 types of 180° side DWs: $SS^+I'$, $FF^+I'$, and $FS^+I'$.

TABLE S3. Strain upon the reference domain after relaxation, DW energies ($E_{DW}$), DW propagation barriers ($\Delta E$) in 1 × 1 × 8 supercells, which consist of two equally spaced domains and two DWs.

| DW type | $E_{DW}$ | | $\Delta E$ | Strain upon the reference domain (%) | | |
|---|---|---|---|---|---|---|
| | ($mJ/m^2$) | ($meV/Å^2$) | (meV) | a | b | c |
| $SS^+1'$ | -38 | -2 | 972 | -0.04 | / | 0.01 |
| $SS^+2'$ | 448 | 28 | 79 | -0.04 | / | 0.23 |
| $SS^+3'$ | 386 | 24 | 693 | -0.48 | / | 0.36 |
| $SS^+4'$ | 399 | 25 | 199 | -0.32 | / | 0.21 |
| $FF^+1'$ | 293 | 18 | 1148 | / | -0.16 | -0.04 |
| $FF^+2'$ | 255 | 16 | 219 | / | 0.30 | 0.31 |
| $FF^+3'$ | 809 | 50 | 186 | / | 0.20 | 0.67 |
| $FF^+4'$ | 506 | 32 | 23 | / | 0.24 | 0.38 |
| $FS^+1'$ | 641 | 40 | 72 | / | 2.09 | 0.22 |
| $FS^+2'$ | 645 | 40 | 105 | / | 2.14 | 0.24 |
| $FS^+3'$ | 641 | 40 | 71 | / | 2.09 | 0.22 |
| $FS^+4'$ | 641 | 40 | 104 | / | 2.09 | 2.22 |

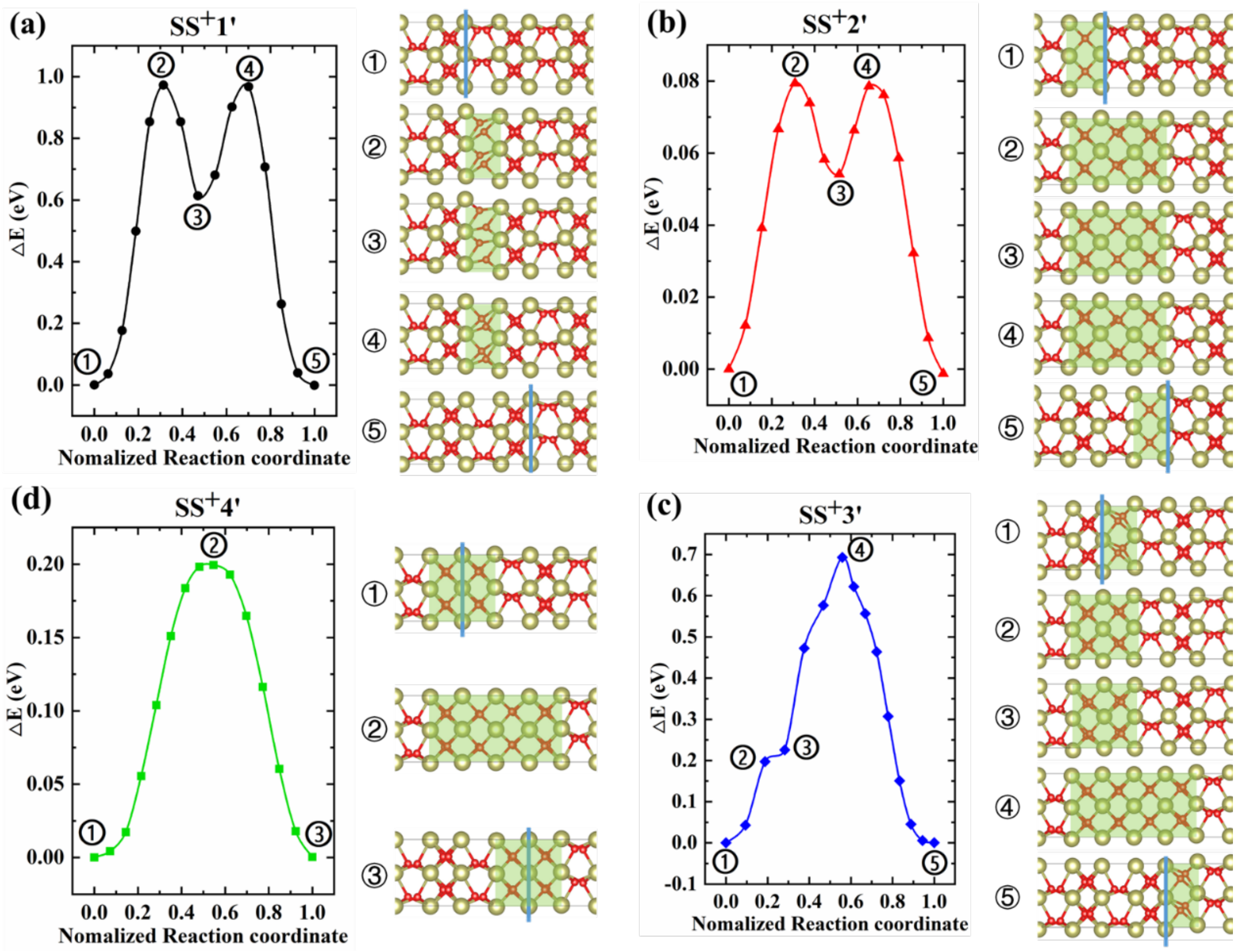

FIG. S11. Details of energy diagram and atomic structure shots of typical images in SS$^{+}$I' type 180° DWs: (a) SS$^{+}$1', (b) SS$^{+}$2', (c) SS$^{+}$3', (d) SS$^{+}$4'. The blue lines indicate roughly the DW location, and the green rectangles mark out distorted area in comparison with normal *Pca*2$_1$-$HfO_2$ unit cell.

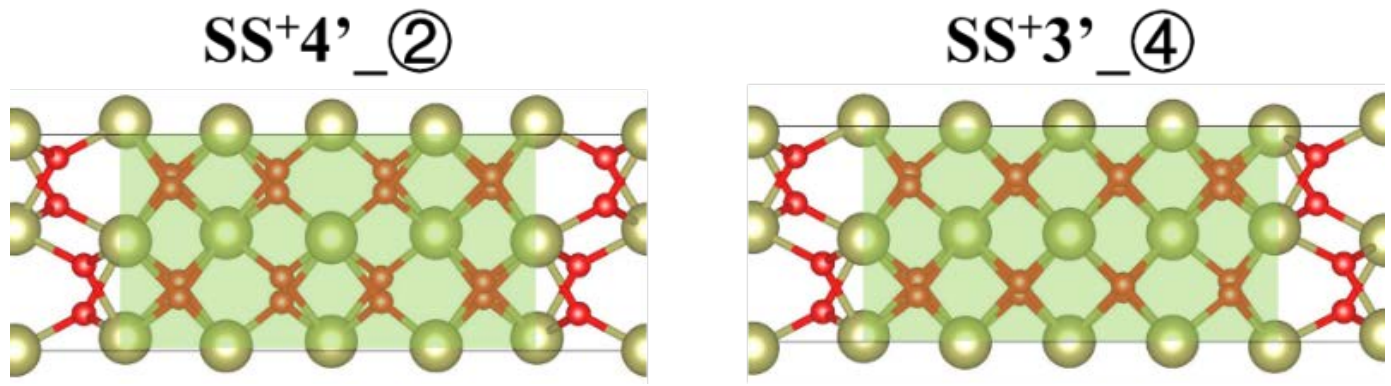

FIG. S12. The top view (from *z*-axis) of two selected DW structures of SS$^{+}$3' and SS$^{+}$4'. They may seem similar from *x*-direction, but are totally different from *z*-direction, and thus a different energy.

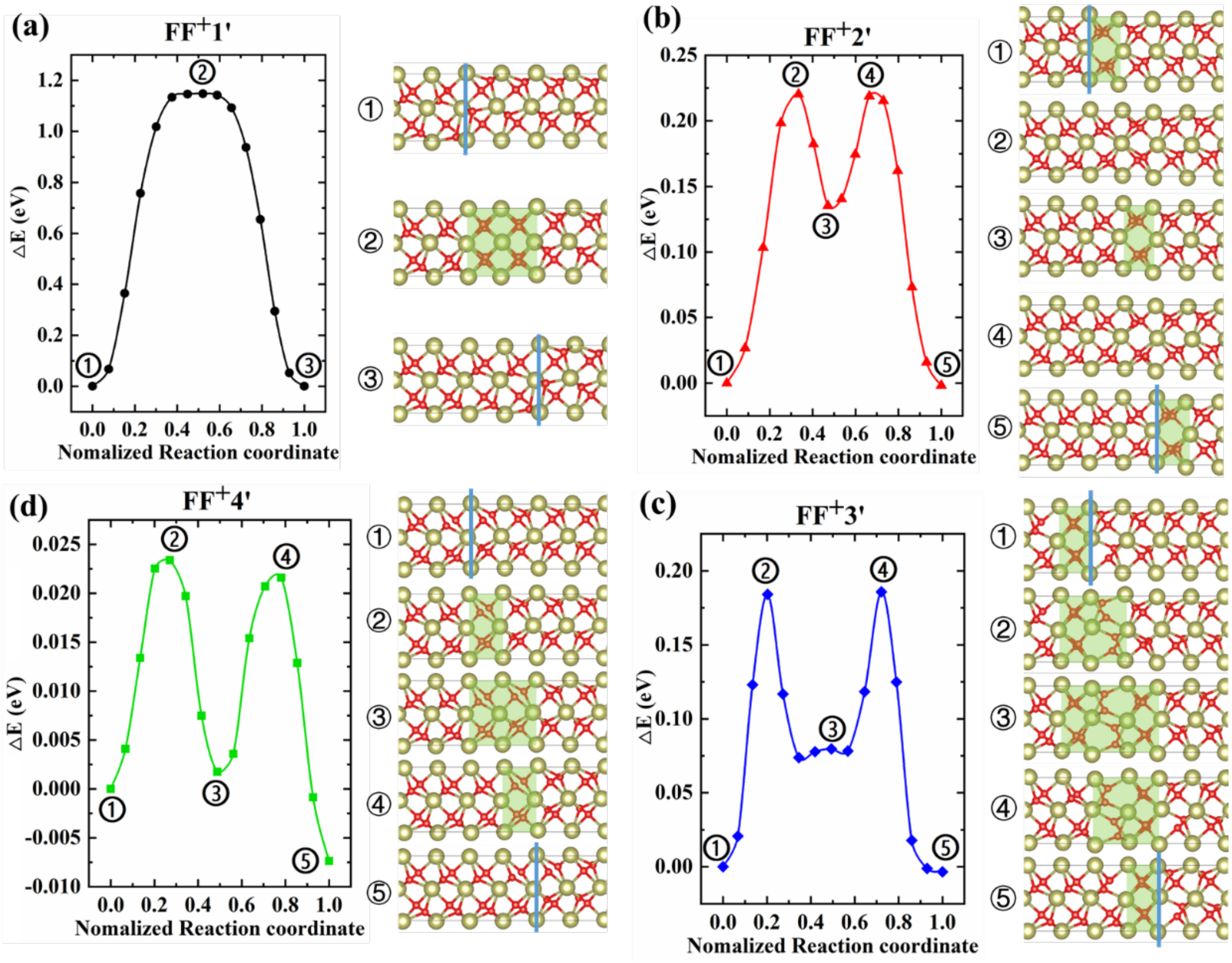

FIG. S13. Details of energy diagram and atomic structure shots of typical images in FF$^{+}$I' type 180° DWs: (a) FF$^{+}$1', (b) FF$^{+}$2', (c) FF$^{+}$3', (d) FF$^{+}$4'. The blue lines roughly indicate the DW location, and the green rectangles mark out distorted area in comparison with normal *Pca*2$_1$-$HfO_2$ unit cell.

Note that our constructed structures for FS$^{+}$1' and FS$^{+}$4' are mutually mixed in the periodic supercells. Since the supercell with two domains unavoidable include these two DWs in the same structure, and this is self-explained in our process of proof for 93 irreducible DW configurations. The situation is the same in FS$^{+}$2' and FS$^{+}$3' DW structures, but there are slight differences in their lattice mismatch and DW energies. The discrepancy is due to the distorted area, or the “DW”, in these two relaxed structures are displaced for half a unit cell, as shown in the right columns of Fig. S14(b, c). This makes one of the domains comparably larger, and brings a larger mismatch energy cost in the already highly strained system.

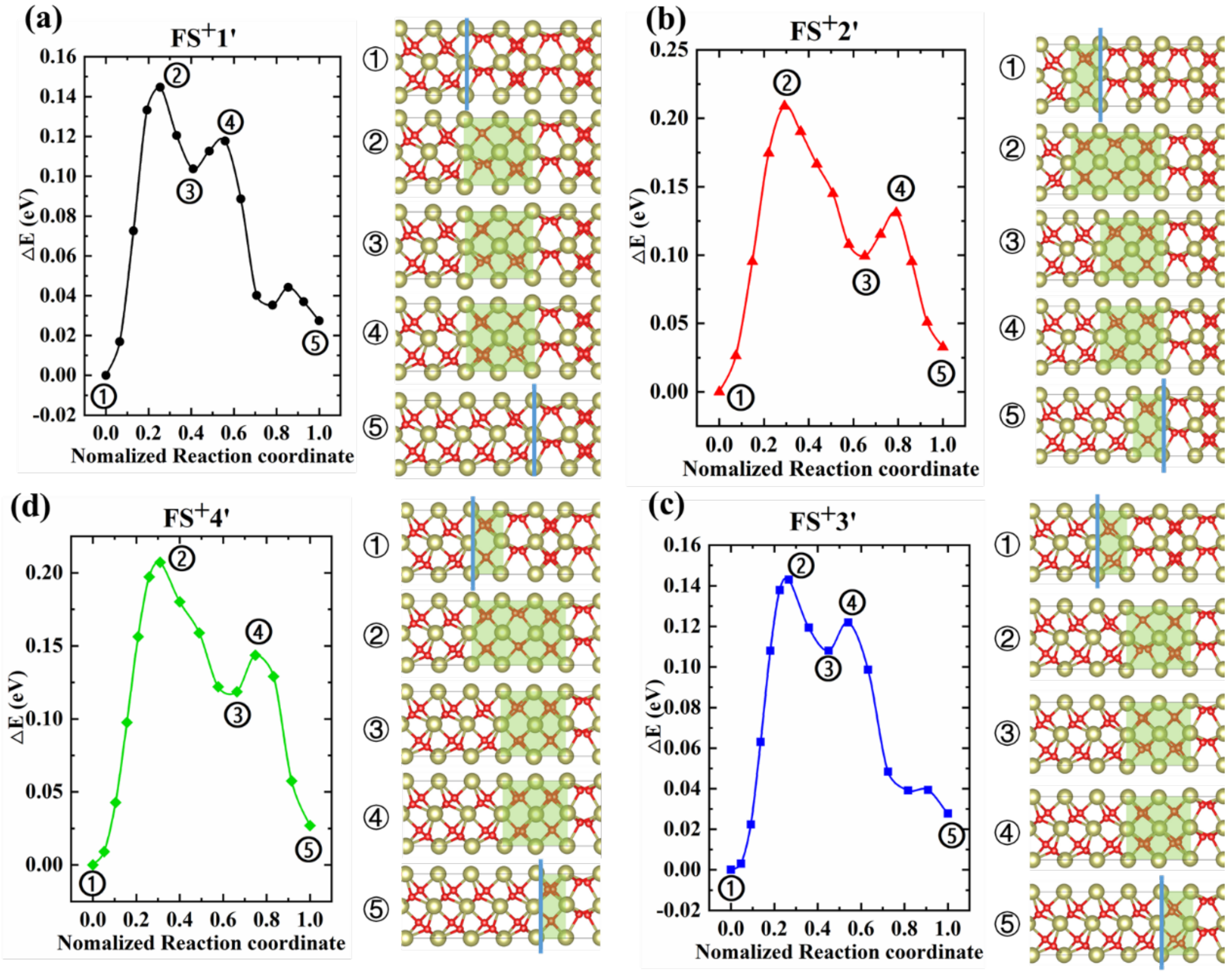

FIG. S14. Details of energy diagram and atomic structure shots of typical images in FS$^{+}$I’ type 180° DWs: (a) FS$^{+}$1’, (b) FS$^{+}$2’, (c) FS$^{+}$3’, (d) FS$^{+}$4’. The blue lines indicate roughly the DW location, and the green rectangles mark out distorted area in comparison with normal *Pca*$2_1$-$HfO_2$ unit cell.

To conclude, in SS$^{+}$I’, FF$^{+}$I’, or FS$^{+}$I’ types of 180° side DWs have varies of DW energies and DW propagation barriers. Thus, to study the mobility of DWs and performance of $HfO_2$-based ferroelectric devices, it is insufficient to consider only the generally studied SS$^{+}$I’ type DWs. $E_{DW}$ ranks are: SS$^{+}$1’ < SS$^{+}$3’ < SS$^{+}$4’ < SS$^{+}$2’, FF$^{+}$1’ < FF$^{+}$2’ < FF$^{+}$4’ < FF$^{+}$3’, and almost stationary among FF$^{+}$I’ type 180° DWs. ΔE ranks are: SS$^{+}$2’ < SS$^{+}$4’ < SS$^{+}$3’ < SS$^{+}$1’, FF$^{+}$4’ < FF$^{+}$3’ < FF$^{+}$2’ < FF$^{+}$1’, and FS$^{+}$3’ ≈ FF$^{+}$1’ < FF$^{+}$4’ ≈ FF$^{+}$2’.

SS$^{+}$I’, FF$^{+}$I’, or FS$^{+}$I’ types of 180° side DWs have varies of DW energies and DW propagation barriers. Thus, to study the mobility of DWs and performance of $HfO_2$-based ferroelectric devices, it is insufficient to consider only the SS$^{+}$I’ type DWs.